# AN 800 MEV SUPERCONDUCTING LINAC TO SUPPORT MEGAWATT PROTON OPERATIONS AT FERMILAB*

Paul Derwent, Steve Holmes and Valeri Lebedev[#], Fermilab, PO box 500, Batavia, IL 60510, USA

*Abstract*

Active discussion on the high energy physics priorities in the US carried out since summer of 2013 resulted in changes in Fermilab plans for future development of the existing accelerator complex. In particular, the scope of Project X was reduced to the support of the Long Base Neutrino Facility (LBNF) at the project first stage. The name of the facility was changed to the PIP-II (Proton Improvement Plan). This new facility is a logical extension of the existing Proton Improvement Plan aimed at doubling average power of the Fermilab's Booster and Main Injector (MI). Its design and required R&D are closely related to the Project X. The paper discusses the goals of this new facility and changes to the Project X linac introduced to support the goals.

## PIP-II DESIGN CRITERIA

A number of approaches based on upgrades to the existing Fermilab accelerator complex can be taken to achieve beam power on the LBNF target in excess of 1 MW. The challenge is to identify a solution that provides an appropriate balance between minimization of near-term costs and flexibility to support longer-term goals. In order to constrain consideration to a modest number of options the following criteria were applied to possible solutions [1]:

- The plan should support the delivery of 1.2 MW from the MI to the LBNF target at energies between 80-120 GeV;
- The plan should provide support to the currently envisioned 8 GeV program, including Mu2e, g-2, and the suite of short baseline neutrino experiments;
- The plan should provide a platform for eventual extension of beam power to LBNF to >2 MW;
- The plan should provide a platform for eventual development of a capability to support multiple rare processes experiments with high duty factor beams, at high beam power.

The primary bottleneck limiting beam power to the LBNF target is related to the existing Linac and Booster. Performance is limited to about $4.4 \times 10^{12}$ protons per Booster pulse by beam loss – primarily driven by the incoherent tune shift due to space-charge at the 400 MeV injection. The secondary bottleneck is the slip-stacking of twelve Booster pulses in the Recycler presently resulting in ~5% beam loss. This loss needs to be reduced for operation at larger beam power.

An ideal facility meeting the above criteria would be the pairing of a new linac with a modern rapid cycling synchrotron capable to accelerate a beam current large enough to avoid slip-stacking. Taking into account the limited acceptance of the Recycler such choice requires an injection energy of about 2 GeV. However cost limitations do not allow us to implement such plan in one step, consequently, requiring a staged approach to the upgrade of the FNAL accelerator complex.

To address the most immediate needs we propose to replace the existing 400 MeV linac by a new 800 MeV super-conducting (SC) linac. To further reduce the cost of the new machine we plan to reuse the existing Tevatron cryogenics infrastructure. Its limited cooling power requires SC linac operation in the pulsed regime. The increased injection energy should allow Booster operation with ~1.7 times larger Booster beam current. To be compatible with CW operation we limit the SC linac beam current to 2 mA. That requires an increase in number of injection turns from 12 to ~300. Although it looks as a quite large increase this number of injection turns is still about 3 times lower than used for injection to the SNS and an analysis shows that it does not present outstanding problems. To further increase the proton flux in the Booster we plan to increase its repetition rate from 15 to 20 Hz. That should also result in a decrease of the beam loss during slip-staking in Recycler. Such a cost effective approach addresses an increase of beam power required by the LBNF and creates a wide range of possibilities for future upgrades. Table 1 presents main parameters of new facility.

Table 1: Main PIP-II parameters

| Parameter | Value | Unit |
| --- | --- | --- |
| Linac beam energy | 800 | MeV |
| Linac beam current | 2 | mA |
| Linac pulse duration | 0.55 | ms |
| Linac/Booster pulse repetition rate | 20 | Hz |
| Linac upgrade potential | CW | |
| Booster Protons per Pulse (extract.) | $6.5 \times 10^{12}$ | |
| Booster Beam Power at 8 GeV | 160 | kW |
| MI Cycle Time @ 120 GeV | 1.2 | s |

## SC LINAC

The linac was described in details in the Project X Reference Design Report (RDR). Here we point out the main features and deviations from it [2, 3]. Figure 1 shows the structure of the linac. A room temperature (RT) section accelerates the beam to 2.1 MeV and creates a desired bunch structure for injection into the superconducting (SC) linac. All accelerating structures are CW compatible. The operation with peak current up to 10 mA is supported by ion source, Low Energy Beam Transport (LEBT) and RFQ. The bunch-by-bunch chopper located in the Medium Energy Beam Transport (MEBT) removes undesired bunches leaving the beam current up to 2 mA (averaged over a few µs period) for further acceleration. In the course of Booster injection the chopper removes bunches at the boundaries of RF buckets and forms a 3-bunch long abort gap in Booster. There is also a "slow" chopper in the LEBT. Its rise and fall times are about 100 ns. It allows one to form a macro-structure in the beam timing required for machine commissioning and to avoid unnecessary beam loading in the MEBT in normal operations. Together the LEBT and MEBT choppers form a desired bunch structure.

___________________________________________

* Work supported by Fermi Research Alliance, LLC, under Contract No. DE-AC02-07CH11359 with the United States Dep. of Energy
[#]val@fnal.gov

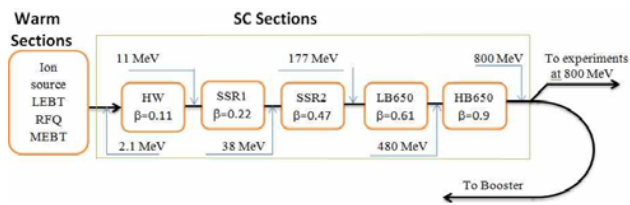

Figure 1: The PIP-II linac structure.

The RFQ energy of 2.1 MeV is chosen because it is below the neutron production threshold for most materials. At the same time this energy is sufficiently large to mitigate the space charge effects in the MEBT at currents as high as 10 mA. The choice of a comparatively low energy for the LEBT (30 keV) allows reducing the length of RFQ adiabatic buncher, and, consequently, achieving sufficiently small longitudinal emittance so that at the exit of the RFQ the beam phase space would be close to the emittance equipartitioning. To mitigate space-charge effects in the LEBT, compensation of beam space charge by residual gas ions can be applied either for the full or partial LEBT length.

Table 2: Main parameters of SC linac cavities

| Name | $\beta_{opt}^{*}$ | Freq. MHz | $B_{peak}$ mT | $E_{peak}$ MV/m | $\Delta E$ (MeV) |
|---|---|---|---|---|---|
| HWR | 0.112 | 162.5 | 41 | 38 | 1.7 |
| SSR1 | 0.222 | 325 | 58 | 38 | 2.05 |
| SSR2 | 0.471 | 325 | 70 | 40 | 4.98 |
| LB650 | 0.647 | 650 | 70 | 37.5 | 11.6 |
| HB650 | 0.950 | 650 | 64 | 35.2 | 17.7 |

* $\beta_{opt}$ is the beam beta where maximum acceleration is achieved, while β presented in Figure 1 for the last two cavity types is the geometric β.

The SC linac starts immediately downstream of the MEBT, accelerating the H⁻ beam from 2.1 to 800 MeV. The operational parameters for the linac SC cavities are presented in Table 2. Five types of superconducting cavities are used to cover the entire velocity range required for beam acceleration. They are presented by:

- One accelerating SC cryomodule based on 162.5 MHz Half-Wave Resonators (HWR) [4,5];
- Two sections of accelerating SC cryomodules based on 325 MHz Single-Spoke Resonators (SSR1 & SSR2);
- Two sections of accelerating SC cryomodules operating at 650 MHz and based on 5-cell elliptical cavities (LB650 and HB650).

Parameters of the cryomodules (CM) are presented in Table 3. The cavity frequencies and cell configurations are chosen to maximize acceleration efficiency for each accelerating structure, minimize the cost of the accelerator and its operation, and to address other factors helping to minimize beam loss. The first 3 types of CMs use the solenoidal focusing with SC solenoids located inside the cryomodules. The periodicity of focusing elements is chosen so that to minimize harmful effects of head-to-tail variations of cavity defocusing fields. That requires a focusing element (solenoid) preceding each cavity in the first cryomodule. With acceleration the periodicity of focusing elements can be relieved. Therefore there are 2 cavities per solenoid in the SSR1 and SSR2 cryomodules. However the solenoids are still located inside cryomodules to minimize focusing period. Focusing in LB650 and HB650 CMs is produced by quadrupole doublets located outside. That significantly reduces cryomodule complexity, and what is more important, removes magnetic field from CMs greatly simplifying magnetic shielding and simplifying an achievement of high value for cavity $Q_0$.

Table 3: Main parameters of SC linac cryomodules

| Name | N CM | Cav./CM | CM* config. | Length (m) |
|---|---|---|---|---|
| HWR | 1 | 8 | 8×(sc) | 5.93 |
| SSR1 | 2 | 8 | 4×(csc) | 5.2 |
| SSR2 | 7 | 5 | sccsccsc | ~6.5 |
| LB650 | 10 | 3 | ccc | ~3.9 |
| HB650 | 4 | 6 | cccccc | ~9.5 |

*c denotes a SC cavity, and s solenoid.

The energy stored in the SC cavities is quite large. That allows one to keep the accelerating voltage fluctuations due to beam loading below $10^{-3}$ if the bunch structure is repetitive with period below about 3 µs.

To support the beam injection to the Booster a pulsed operation of the linac is sufficient. In this case the linac operates at 20 Hz with beam pulse duration of 0.55 ms resulting in 1.1% beam duty factor. Cavity filling with RF requires significantly longer time (see Figure 2). The effective duty factor for cryogenic load is about 6.6%. The effective duty factor for high power RF is about 13%. To reduce cryogenic power the phase of RF amplifiers can be shifted by 180° to accelerate voltage decay in a cavity after the beam pulse.

The RF system is based on a single RF source driving each cavity, for a total of 114 RF sources. It is anticipated that the amplifiers in the 162.5 and 350 MHz sections will be solid state, while those in the 650 MHz sections will be either inductive output tubes (IOTs) or solid state. We also consider a possibility to use a phase locked magnetrons [6] as power amplifiers.

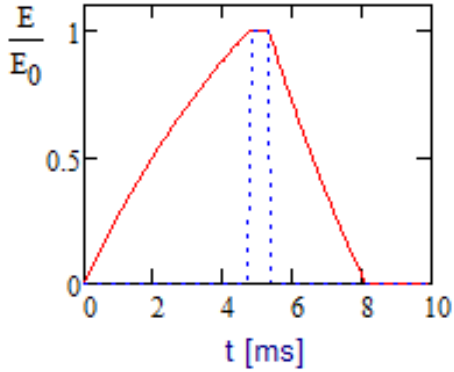

Figure 2: Dependence of RF voltage on time for pulsed linac operation. Blue dashed line shows the beam pulse.

The average RF power delivered to the cavities consists of two contributions: the energy transferred to the beam; and the energy required to fill and discharge the accelerating cavities. The second contribution is about 10 times larger than the first and, in general, the average power associated with this contribution does not depend on the peak power of RF amplifier. For a fixed average power the RF cost increases with peak power and therefore the RF cost minimum is achieved with RF power equal to that required to accelerate the beam. One consequence of this strategy is that the cost savings associated with the pulsed power amplifiers in going from CW to low duty factor is modest (~10%) and therefore CW capable RF amplifiers are planned. The RF requirements are summarized in Table 4. The presented powers also include power margins required to control microphonics and the Lorentz force detuning. The latter is expected to be a serious challenge. To keep a cavity at resonance in addition to a slow mechanical tuner we will be using a fast piezo-tuner. It will also require state of the art microphonics and low level RF controls.

Table 4: Cavity bandwidths and required RF power

| Name | Maximal detune (Hz) | Minimal half-bandwidth, $f_0/2Q$ (Hz) | Maximum required power (kW) |
|---|---|---|---|
| HWR | 20 | 34 | 4.8 |
| SSR1 | 20 | 45 | 5.3 |
| SSR2 | 20 | 27 | 17 |
| LB650 | 20 | 29 | 33 |
| HB650 | 20 | 31 | 49 |

The required power of the cryogenic system is determined by static and dynamic loads. In estimates of required power we use a conservative approach for $Q_0$ values of SC cavities. We assume the following values of $Q_0$: $5\times10^9$ for HWR and SSR1, $1.2\times10^{10}$ for SSR2, $1.5\times10^{10}$ for LB650, and $2\times10^{10}$ for HB650. As one can see from Table 5 the dynamic load is significantly lower for the pulsed regime than for CW. However the dynamic power strongly dominates in CW regime. Recent successes in our $Q_0$ program [7] are extremely encouraging and suggest that a $Q_0$ increase by more than factor of 2 with approximately the same reduction of required cryogenic power is achievable. Fermilab is pursuing an intense R&D program to transfer $Q_0$ values achieved in vertical tests to cavities operating in a real cryomodule [8].

To minimize the cost of the PIP-II cryogenic system it will be assembled utilizing considerable existing Tevatron cryogenic infrastructure, including the Central Helium Liquefier (CHL), transfer line, and compressors. The cryo-plant cooling power is: 5729 W at 70K, 1250 W at 5K and 490 W at 2K. As one can see there is sufficient margin at all temperatures. A future upgrade to CW operation would require a new 2K cryogenic plant even if mentioned above $Q_0$ values will be achieved.

Table 5: Requirements to the cryogenic power

| Name | Static load per CM (W) | | | Dynamic load per CM (W) | |
|---|---|---|---|---|---|
| | 70K | 5K | 2K | 2K CW | 2K Pulsed |
| HWR | 250 | 60 | 14 | 10 | 10* |
| SSR1 | 195 | 70 | 16 | 11 | 11* |
| SSR2 | 145 | 50 | 8.8 | 43 | 2.8 |
| LB650 | 85 | 25 | 5 | 73 | 4.8 |
| HB650 | 120 | 30 | 6.2 | 147 | 9.7 |
| **Total** | **2985** | **920** | **182** | **1651** | **138** |

* These cryomodules operate in CW

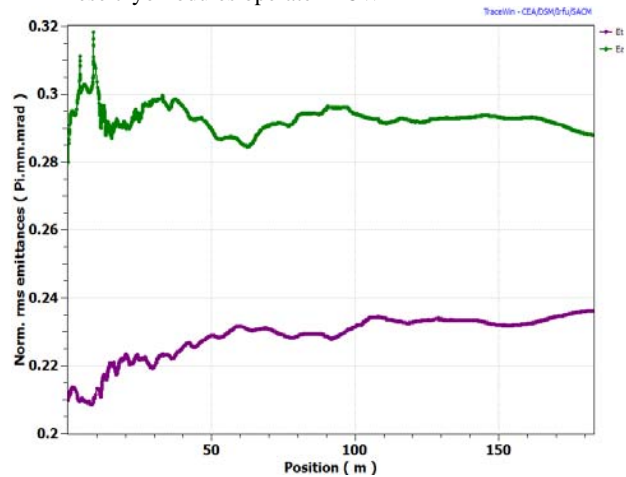

Figure 3: Simulations of longitudinal (top curve) and transverse (bottom curve) emittance evolution in the PIP-II linac; beam current - 5 mA.

The beam dynamics in the PIP-II linac is very similar to that in the Project X linac. Figure 3 presents simulations of the emittance growth in the course of the acceleration. As one can see a moderate emittance growth is found. The final values of the emittance are within the PIP-II specifications.

## RECENT PROGRESS

Fermilab is carrying out an extensive R&D program in support of PIP-II. Presently, it is mainly focused on the design and construction of PXIE [9] which consists of the normal conducting linac frontend and first two SC cryomodules. We expect to have PXIE operating in 2018. R&D on other cryomodules was also recently initiated.

The LEBT [10] has been recently installed in the PXIE enclosure and its beam commissioning is underway. The RFQ is being built by LBNL and is expected to be delivered in the spring of 2015. Designs of HWR and SSR1 cryomodules are complete and production started. Their delivery is expected in 2017.

## CONCLUSIONS

After years of discussions and studies Fermilab formulated the path for the upgrade of its accelerator complex. The project is presently in its initial phase and is expected to proceed expeditiously with the main goal to achieve the beam power at the LBNF in excess of 1 MW by 2024.

## ACKNOWLEDGEMENTS

The authors are grateful to our colleagues from Fermilab, LBNL, ANL, SNS, IUAC, BARC, RRCAT, and VECC which work on the PIP-II project. Writing this paper would not be possible without their contributions. In particular we would like to thank: T. Nicol, A. Saini, A. Shemyakin and V. Yakovlev for help and advice in preparing this paper.